\documentclass[prodmode,acmtos]{acmsmall}

\usepackage{url,amsmath,amssymb,epstopdf,framed,multirow,array}

\usepackage[ruled]{algorithm2e}

\SetAlFnt{\small}
\SetAlCapFnt{\small}
\SetAlCapNameFnt{\small}
\SetAlCapHSkip{0pt}
\IncMargin{-\parindent}

\acmVolume{0}
\acmNumber{0}
\acmArticle{0}
\acmYear{2012}
\acmMonth{0}

\begin{document}

\markboth{T. Frankie et al.}{Analysis of Trim Commands on Overprovisioning and Write Amplification in Solid State Drives}

\title{Analysis of Trim Commands on Overprovisioning and Write Amplification in Solid State Drives}
\author{TASHA FRANKIE
\affil{ECE Dept., UC San Diego}
GORDON HUGHES
\affil{CMRR, UC San Diego}
KEN KREUTZ-DELGADO
\affil{ECE Dept., UC San Diego}
}

\begin{abstract}
This paper presents a performance model of the ATA/ATAPI SSD Trim command under various types of user workloads, including a uniform random workload, a workload with hot and cold data, and a workload with $N$ temperatures of data. We first examine the Trim-modified uniform random workload to predict utilization, then use this result to compute the resultant level of effective overprovisioning. This allows modification of models previously suggested to predict write amplification of a non-Trim uniform random workload under greedy garbage collection.  Finally, we expand the theory to cover a workload consisting of hot and cold data (and also $N$ temperatures of data), providing formulas to predict write amplification in these scenarios.
\end{abstract}





\acmformat{Frankie, T., Hughes, G., Kreutz-Delgado, K. 2012. Analysis of Trim Commands on Overprovisioning and Write Amplification in Solid State Drives.}

\begin{bottomstuff}

Author's addresses: T. Frankie {and} K. Kreutz-Delgado, Electrical and Computer Engineering Department, University of California, San Diego; G. Hughes, Center for Magnetic Recording Research, UC San Diego.
\end{bottomstuff}

\maketitle

\section{Introduction}
NAND flash nonvolatile memory has become ubiquitous in electronic devices.  In its form as a solid-state drive (SSD), it is often used as a replacement for the traditional hard disk drive, due to its faster performance.  However, flash devices have some drawbacks, notably a performance slowdown when the device is filled with data, and a limited number of program/erase cycles before device failure.

The reason flash-based SSDs slow as they fill with data is write amplification, in which more writes are performed on the device than are requested.  This is due to the dynamic nature of the logical-to-physical mapping of data in conjunction with the erase-before-write requirement of NAND flash.  A more detailed explanation of write amplification and its causes is in Section~\ref{sec:NAND_primer}.

Manufacturers use several techniques to reduce write amplification in an effort to mitigate the negative performance impact.  Two of these techniques are physically overprovisioning the device with extra storage, and implementing the Trim command~\cite{TrimSpec}.  More information about these techniques is in section~\ref{sec:NAND_primer}.

The main contributions of this work can be summarized as follows.
\begin{itemize}
\item A Markovian birth-death model of random workloads, previously described by ~\citeN{frankie2012_acmse} is shown, along with a new analysis of higher order terms.
\item The write amplification of the Trim-modified random workload, previously described by ~\citeN{frankie2012_ciit}, is extended to hot and cold data.
\end{itemize}

\section{A NAND Flash Primer}\label{sec:NAND_primer}
Some familiarity with NAND flash is required to understand the models presented in this paper.  In this section, we summarize the necessary NAND flash background material for the reader.

\subsection{Flash Layout}
NAND flash is composed of pages, which are grouped together in physical blocks\footnote{Note that physical blocks are different than logical blocks, and the type referred to when the word ``block" is used needs to be determined from the context.  In this paper, we attempt to specify logical or physical block if the context is potentially unclear.}.  Pages are the smallest write unit, and hold a fixed amount of data, the size of which varies by manufacturer, and is typically 2K, 4K, or 8K in size~\cite{grupp2009characterizing,roberts2009integrating}.  Unlike with traditional magnetic media, pages cannot be directly overwritten, but must first be erased before they can be programmed again, a requirement dubbed erase-before-write.  Blocks are the smallest erase unit, and typically contain 64, 128, or 256 pages, again varying by manufacturer~\cite{grupp2009characterizing}.

Because of the erase-before-write requirement and the mismatch in sizes between the smallest write unit (pages) and the smallest erase unit (blocks), write-in-place schemes are impractical in flash devices.  To overcome this issue, most flash devices employ a dynamic logical-to-physical mapping called the Flash Translation Layer (FTL)~\cite{chung2009survey,gupta2009dftl}.  There are many FTLs discussed in the literature; a survey and taxonomy of these is found in~\cite{chung2009survey}.  In this paper, we employ a log-structured SSD, as described by~\citeN{Hu2009}, in which the data associated with a write request is written to the next available pages.  If this data was intended to overwrite previously recorded data, the pages on which the data was previously stored are invalidated in the FTL mapping.

\subsection{Garbage Collection}
At some point in the process of writing data, most of the pages will be filled with either valid or invalidated data, and more space is needed for writing fresh data. Garbage collection occurs in order to reclaim additional blocks.  In this process, a victim block is chosen for reclamation, its valid pages are copied elsewhere to avoid data loss, then the entire block is erased and the FTL map pointer is changed.  Many algorithms exist for choosing the victim block, most of which are based on either selecting the block that creates the most free space or one that also takes into account the age of the data on the block, as originally described in the seminal paper by~\citeN{rosenblum1992design}.  In this paper, we utilize the greedy garbage collection method: when the number of free (erased) blocks in the reserve-queue drops below a predetermined threshold, $r$, the block with the fewest number of valid pages is chosen for reclamation, as explained in Algorithm~\ref{alg:greedy_garbage_collection}.

\begin{algorithm}[t]
\SetAlgoNoLine
\KwIn{LBA requests of types Write and Trim.}
\KwOut{Data placement and garbage collection procedures.}

\While{there are LBA requests}{
      \eIf{request is Write}{
        write new data to first free page in current-block\;
        \If{LBA written was previously In-Use}{
          invalidate appropriate page in FTL mapping\;
        }
        \If{size(free pages in current-block)$=0$}{
          append current-block to end of occupied-block-queue\;
          current-block $\leftarrow$ first block in reserve-queue\;
          \If{size(reserve-queue)$<r$}{
            \tcp{garbage collection:}
            victim-block $\leftarrow$ block with fewest valid pages from occupied-block-queue\;
            copy valid pages from victim-block to current-block\;
            erase victim-block and append to reserve-queue\;
          }
        }
      }{
      \tcp{request was Trim}
      invalidate appropriate page in FTL mapping\;
      }

}
\caption{Data Placement and Greedy Garbage Collection}
\label{alg:greedy_garbage_collection}
\end{algorithm}

\subsection{Write Amplification}
During garbage collection, extra page writes not directly requested of the SSD occur.  These extra writes contribute to a phenomenon known as write amplification.  Write amplification, measured as the ratio of the average number of actual writes to requested writes~\cite{Hu2009}, slows the overall write speed of the device as it fills with data and reduces the lifetime of the device, since each page has a limited number of program/erase cycles before it wears out and can no longer be used to store data.

Most FTLs and garbage collection algorithms attempt to reduce write amplification and perform wear leveling, attempting to keep the number of program/erase cycles uniform over all blocks.  Greedy garbage collection has been proven optimal for random workloads in terms of write amplification reduction~\cite{hu2010fundamental}.

\subsubsection{Overprovisioning}
In addition to FTLs and garbage collection methods that attempt to work well with various workload, manufacturers use other ways to reduce write amplification.  One of these methods is through physically overprovisioning the flash device by providing more physical storage than the user is allowed to logically access.  Overprovisioning reduces write amplification by increasing the amount of time between garbage collections of the same block, which in turn increases the probability that more pages in the block are invalid at the time of garbage collection, and thus need not be copied.  The mathematical relationship between overprovisioning and write amplification in uniform random workloads has been studied by~\citeN{Hu2009} and~\citeN{hu2010fundamental},~\citeN{agarwal2010closed}, and~\citeN{xiang2011improved}.

In this paper, we measure the level of overprovisioning using the spare factor: $S_f = (t-u)/t$.  The spare factor has a range of [0,1], making comparisons of various spare factors intuitive.

\subsubsection{Trim}
Another way that write amplification can be reduced is through use of the Trim command, defined in the working draft of the ATA/ATAPI Command Set - 2 (ACS-2)~\cite{TrimSpec}.  The Trim command allows the file system to communicate to the SSD that data is no longer needed, and can be treated as invalidated.  This means that the data in these pages does not need to be retained during garbage collection, reducing the number of extra writes that garbage collection generates.  As described by~\citeN{frankie2012_acmse}, and detailed in section~\ref{sec:trim_model_markov}, the Trim command effectively increases the amount of overprovisioning on the SSD.

\section{Workloads}
There are many different workloads seen in the real world, from a single user of a personal computer, to multiple users of cloud computing, database transactions, sequential serial access, and random page access.  However, workloads can be seen as existing along a spectrum from purely sequential requests to completely random requests, with most real-world workloads falling between the two extremes.  Purely sequential workloads are not very interesting to study in conjunction with write amplification in SSDs, because by nature of being sequential, entire blocks are invalidated as requests continue.  This means that at garbage collection, there is no valid data that needs to be saved in the pages of the block being reclaimed, and so there are no extra writes to contribute to write amplification~\cite{Hu2009}.  On the other hand, random workloads cause considerable write amplification; the effect on write amplification of uniform random workloads without Trim has been studied by~\citeN{Hu2009} and~\citeN{hu2010fundamental},~\citeN{agarwal2010closed}, and~\citeN{xiang2011improved}.  The effect on write amplification of uniform random workloads with Trim has been analyzed by~\citeN{frankie2012_ciit}.

For the study of write amplification, read requests are irrelevant, and so are not considered in the uniform random workload.  The standard uniform random workload studied by~\citeN{Hu2009} and~\citeN{hu2010fundamental},~\citeN{agarwal2010closed}, and~\citeN{xiang2011improved} consists only of Write requests that are uniformly randomly chosen over all $u$ user logical block addresses (LBAs).  To simplify the math involved in the analysis, it is assumed that the data associated with a write request to one LBA will occupy one physical page.  \citeN{frankie2012_ciit} introduced Trim requests to the standard uniform random workload.  In this Trim-modified uniform random workload, requests are split between Writes and Trims, with Trims occurring with probability $q$.  Write requests are uniformly random over all $u$ user LBAs, and Trim requests are uniformly random over all In-Use LBAs.  \citeN{frankie2012_ciit} define an In-Use LBA as one whose most recent request was a Write; all other LBAs, including those whose most recent request was a Trim, or that had never been written, are considered not In-Use.

\section{A Model for Trim}\label{sec:trim_model_markov}
We have previously shown that a Trim-modified uniform random workload can be modeled as a Markov birth-death chain~\cite{frankie2012_acmse}.  We summarize the results here, showing the necessary details in order to also perform a higher-order analysis.  For the convenience of the reader, a list of variables and their descriptions used in the paper is given in Table~\ref{table:var_descrip}.

\begin{table}
\tbl{List of variables and their descriptions\label{table:var_descrip}}{
\begin{tabular}{|c|l|} \hline
Variable&Description\\ \hline
$t$ & Number of pages on SSD\\ \hline
$u$   & Number of pages user is allowed to fill\\ \hline
$n_p$ & Number of pages in a block\\ \hline
$r$ & Minimum number of reserved blocks in reserve-queue\\ \hline
$S_f=(t-u)/t$ & Manufacturer-specified spare factor\\ \hline
$\rho = (t-u)/u$ & Alternate measure of overprovisioning\\ \hline
$q$   & Probability of Trim request\\ \hline
$\pi_x$ & \textit{Unnormalized} steady-state probabilities\\ \hline
$s=(1-2q)/(1-q)$ & Steady-state probability of a Write being for an In-Use LBA\\ \hline
$\bar{s}=q/(1-q)$ & Steady-state probability of a Write being for a not In-Use LBA\\ \hline
$S_\text{eff}=(t-X_n)/t$ & Effective spare factor \\ \hline
$\rho_\text{eff} = \frac{\bar{S}_\text{eff}}{1-\bar{S}_\text{eff}} = \frac{1+\rho}{s}-1$ & Alternate measure of effective overprovisioning \\ \hline
$u_\text{eff} = us$ & Mean number of In-Use LBAs at steady-state\\ \hline
$\sigma^2 = u\bar{s}$ & Variance of the number of In-Use LBAs at steady-state\\ \hline
$f_h, f_c$ & Fraction of the data that is hot or cold\\ \hline
$p_h, p_c$ & Probability of requesting hot or cold data\\
\hline\end{tabular}}

\end{table}


\subsection{Trim-Modified Uniform Random Workload as a Markov Birth-Death Chain}
The Trim-modified uniform random workload is modeled as a Markov birth-death chain in which the number of In-Use LBAs at time $n$ is the state, $X_n$.  Write requests can either increase the state by 1 or leave it unchanged, and Trim requests reduce the state by 1.  The problem-specific transition probabilities of the Markov chain are as follows:

\begin{align*}
P(X_{n+1}=x-1|X_n=x) &= q_x = q\\
P(X_{n+1}=x|X_n=x) &= r_x = \left( \frac{x}{u} \right)\left( 1 - q \right)\\
P(X_{n+1}=x+1|X_n=x) &= p_x = \left( \frac{u - x}{u} \right) \left( 1 - q \right)
\end{align*}
where $q_x$ is the probability of a Trim request, $r_x$ is the probability of a Write request for an In-Use LBA, and $p_x$ is the probability of a Write request for a not In-Use LBA.  Note that the value of $q_x$ is a constant $q$ regardless of the state, but $r_x$ and $p_x$ are dependent on the number of In-Use LBAs.

The steady-state occupation of a Markov birth-death chain is known to be~\cite{hoel1971}
\begin{align*}
  \pi _x \stackrel{\text{def}}{=} P_\text{unnormalized}(X_\text{steady}=x) = \begin{cases}
  \frac{p_0 \cdots p_{x - 1}}{q_1 \cdots q_x}, & x \geqslant 1 \\
  1 & x = 0
  \end{cases}
\end{align*}
Substituting for the problem-specific values, we find
\begin{equation}\label{eqn:ugly_SS}
\pi _x = \left( \frac{1 - q}{q} \right)^x\frac{u!}{u^x\left( u - x \right)!} \, , \qquad \forall x \in \{ 0,...,u\}
\end{equation}
From an inspection of Equation~\ref{eqn:ugly_SS}, it is not straightforward to understand the effect of the values of $u$ and $q$ on the distribution.  However, this equation can be manipulated into a more comprehensible form, as shown in the next subsection.

\subsection{Understanding the Steady-State Occupation Probabilities}
In this section, we manipulate Equation~\ref{eqn:ugly_SS} into a more familiar form, while maintaining asymptotic equivalence to the normalization factor (denoted with $\sim$).  First, define $s = \left( \frac{1 - 2q}{1 - q} \right)$ and $\bar s = 1 - s = \left( \frac{q}{1 - q} \right)$, noting that $s+\bar{s}=1$.  With these substitutions, Equation~\ref{eqn:ugly_SS} becomes
\begin{equation*}
\pi _x = \bar s^{-x}\frac{u!}{u^x\left( {u - x} \right)!}
\end{equation*}
The exponential terms suggest that working in log-space may be easier:
\begin{align*}\label{}
\log \left( {{\pi _x}} \right) &\sim  - \log \left( {\left( {u - x} \right)!} \right) - x\log \left( {u\bar s} \right)
\end{align*}
At this point, it is useful to apply Stirling's formula for factorials: $\log \left( {n!} \right)\approx n\log \left( n \right) - n$, using $\approx$ to denote an asymptotic approximation. It is well known that Stirling's approximation for factorials converges very quickly, and for relatively small values of $n$ can practically be considered to provide a ``ground truth" equivalent for the value of the factorial function\footnote{Stirling's Formula, one of several possible Stirling approximations to the factorial function, provides a highly accurate approximation to $n!$ for values of $n$ as small as 10-to-100.  For this reason, for even moderately sized values of $n$, the Stirling Formula is used to derive a variety of useful probability distributions in fields such as probability theory~\cite{feller1968}, equilibrium statistical mechanics~\cite{reif1965}, and information theory~\cite{cover2006elements}.  Because of the remarkable accuracy of the Stirling Approximation, which very rapidly increases with the size of $n$, the resulting distributions can be taken to be "ground truth" distributions in many practical domains.  Derivations of the Stirling Approximation, and discussions of its accuracy and convergence, can be found in pages 52-54 of~\cite{feller1968}, Appendix A.6 of~\cite{reif1965}, and pages 522-524 of~\cite{courant1953}.}.  The application of Stirling's approximation to this problem is reasonable.  For SSDs of several GB, $u$ is on the order of hundreds of thousands, so the Stirling approximation of $(u-x)!$ is effectively ground truth except for a few $x$ very close in value to $u$.
\begin{align}
 \log \left( {{\pi _x}} \right) &\approx  - \left( {u - x} \right)\log \left( {u - x} \right) + \left( {u - x} \right) - x\log \left( {u\bar s} \right)\nonumber\\
 &\sim  - \left( {u - x} \right)\log \left( {\frac{{u - x}}{{u\bar s}}} \right) + \left( {u - x} \right)\label{eqn:Stirling_approx}
\end{align}

A Taylor series expansion is a natural way to handle the $\log$ term.  First the definition $\sigma^2 = u \bar{s}$ and change of variables
\begin{equation*}
\alpha = \frac{x - us}{u\bar{s}}
\end{equation*}
helps to put the $\log$ term into standard form to apply the Taylor series expansion
\begin{equation*}
 \log(1-\alpha) = -
\sum_{n=1}^\infty \frac{\alpha^n}{n} \ \ \text{for} \ \  -1 \le \alpha
< 1
\end{equation*}
Then,
\begin{align}
\log \left( {{\pi _x}} \right)  &\approx -\sigma^2(1-\alpha)\log (1-\alpha)+\sigma^2(1-\alpha) \nonumber \\
&=\sigma^2(1-\alpha)\sum_{n=1}^\infty \frac{\alpha^n}{n}+\sigma^2(1-\alpha) \nonumber \\
&=-\sigma^2 \left( - \alpha +  \sum_{n=1}^\infty
\frac{\alpha^{n+1}}{n(n+1)} \right) + \sigma^2(1-\alpha)\nonumber \\
&\approx -\sigma^2\left(-\alpha+\frac{\alpha^2}{2}\right) + \sigma^2(1-\alpha) \nonumber \\
&\sim \frac{\alpha^2}{2}  \nonumber \\
&= \frac{-(x-us)^2}{2u\bar{s}} \label{eqn:Gaussian_approx}
\end{align}
and we see that the Taylor series first-order approximation of $\pi_x$ is an unnormalized Gaussian random variable with mean $us$ and variance $u\bar{s}$.\footnote{For proof of convergence of the Taylor series within the range needed for the problem, please refer to ~\cite{frankie2012_acmse}.}

\subsection{Higher Order Terms}
The analysis above allows us to compute higher-order moments such as skewness and kurtosis by truncating the Taylor series expansion at later terms.  The skewness is calculated as\footnote{$R\left(\frac{1}{\sigma^3 }\right)$ is defined to mean that the remaining terms are at least as small as $O\left(\frac{1}{\sigma^3 }\right)$, in terms of the ``big O'' terminology.}
\begin{align}
\text{Skew$_X$} = -  \frac{1}{\sigma}  + R\left(\frac{1}{\sigma^3}\right)
\end{align}
and the kurtosis is calculated as
\begin{align}
\text{Kurtosis$_X$} = \frac{3}{4\, \sigma^2} + R\left(\frac{1}{\sigma^4}\right)
\end{align}
For the details of the derivation of the skewness and kurtosis, see~\cite{TashaDissertation}.

The slight negative skew is toward a higher level of effective overprovisioning.  This is more evident for small $u$ with a narrow variance (low but non-zero $q$).  Fig.~\ref{fig:skew_demo} shows simulation, exact analytic values, and approximate Gaussian values for very small $u=25$.  With this small of a $u$, the slight negative skew is evident.  It is unlikely that such a low value of $u$ would every be encountered in a real device; the higher-order analysis is included here for completeness.

A more detailed asymptotic analysis given in~\cite{TashaDissertation}, shows that even in the limit of large $u$ there is an irreducible, residual  absolute  error in the location of the mean of 0.5 which, consistent with the theory of asymptotic analysis,  is asymptotically relatively negligible $\lim_{u \to \infty} 0.5/u = 0$. We correct for this known residual error in the approximation models via a shift, which results in the shown fits in Fig.~\ref{fig:StirlingGroundTruth} and Fig.~\ref{fig:skew_demo}.

\subsection{Effective Overprovisioning from Trim}
We measure the level of effective overprovisioning through the effective spare factor, defined as
\begin{equation*}
S_{\text{eff}} = \frac{t-X_n}{t},
\end{equation*}
The mean and variance of the effective spare factor are calculated as
\begin{eqnarray*}
\bar{S}_{\text{ eff}} &=& \frac{t-us}{t}\\
&=& \bar s + s S_f
\end{eqnarray*}
and
\begin{eqnarray*}
\text{Var}(S_\text{eff}) &=& \frac{1}{t^2} \text{Var}(X_n) \\
&=& \frac{u\bar{s}}{t^2} \\
&=& \frac{\bar{s}(1-S_f)}{t}
\end{eqnarray*}
Note that the mean effective spare factor is dependent only on the manufacturer-specified spare factor and the amount of Trim in the workload.  However, the variance of the effective spare factor also depends on the absolute size of the device, and tends toward zero as the device size gets large.  This means that for large enough SSDs, the penalty for ignoring the variance of the effective spare factor is negligible.

\subsection{Simulation Results and Practical Application}
\begin{figure}
\centerline{\includegraphics{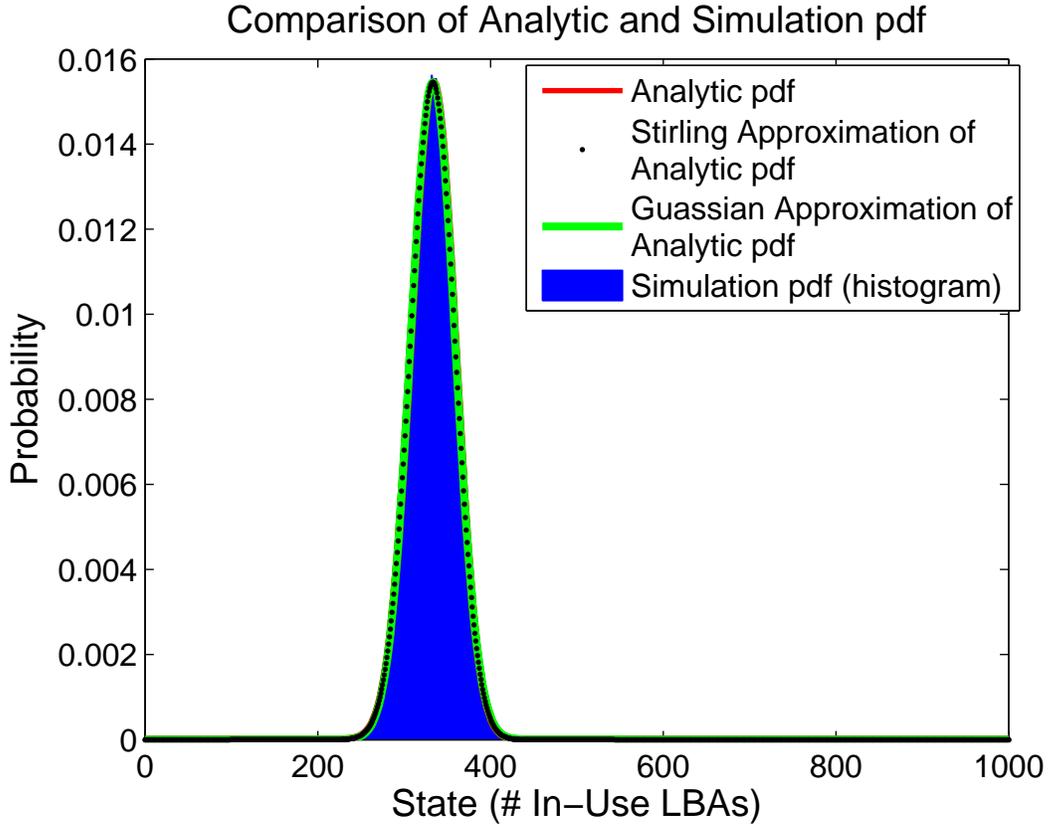}}
\caption{A histogram of one run of simulation results for $u=1000$ and $q=0.4$, along with plots of the exact analytic pdf (Equation~\ref{eqn:ugly_SS}) in red, the Stirling approximation of the pdf (Equation~\ref{eqn:Stirling_approx}) in black dots, and the Gaussian approximation of the pdf (Equation~\ref{eqn:Gaussian_approx}) in green.  The line for the exact analytic pdf is not even visible because the line for the Gaussian approximation of the pdf covers it completely.  The Stirling approximation of the pdf (black dots) lines up well with the Gaussian approximation and the (covered) analytic pdf.  Even at this low value of $u=1000$, the theoretical approximations are effectively identical to the exact analytic pdf.}
\label{fig:StirlingGroundTruth}
\end{figure}

\begin{figure}
\centerline{\includegraphics{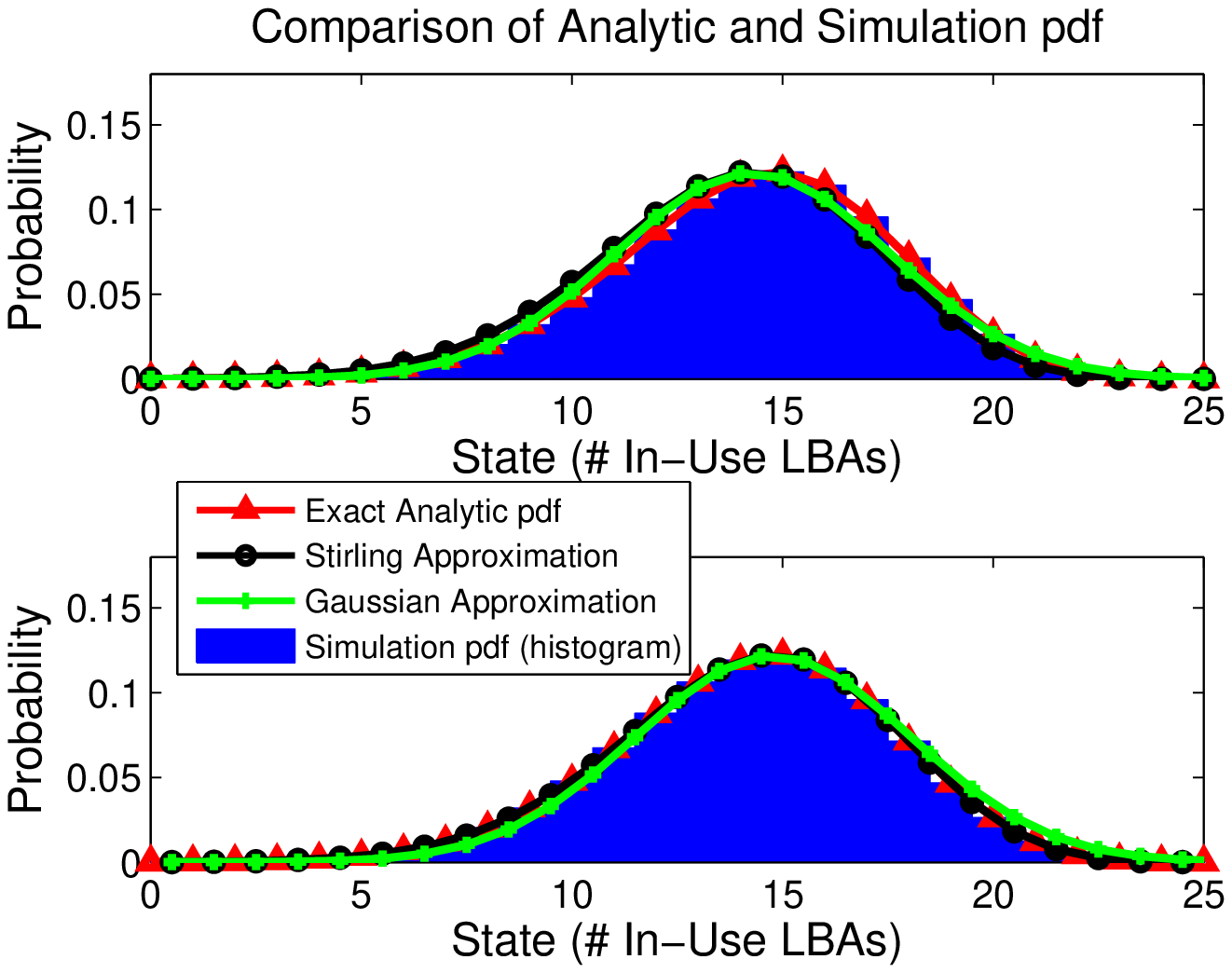}}
\caption{A histogram of one run of simulation results for $u=25$ and $q=0.3$, along with plots of the exact analytic pdf (Equation~\ref{eqn:ugly_SS}) in red, the Stirling approximation of the pdf (Equation~\ref{eqn:Stirling_approx}) in black, and the Gaussian approximation of the pdf (Equation~\ref{eqn:Gaussian_approx}) in green.  A slight negative skew of the exact analytic pdf is evident when compared to the Gaussian approximation.  Note that the exact analytic pdf is the best fit of the histogram.  A series of 64 Monte-Carlo simulations with these values result in an average skew of mean $\pm$ one standard deviation $=-0.299 \pm 0.004$, which agrees well with the theoretically computed approximation value of $-0.306 \pm R\left(\frac{1}{\sigma^3}\right) = -0.306 \pm  R(0.029)$.  The average simulation kurtosis is 0.064; the approximate theoretical is 0.070, both values well within overlapping error bounds.  The top graph is a plot of the numerically normalized equations as shown in the text.  In the bottom graph, a half-bin shift correction has been made to the Stirling and Gaussian approximation densities as described in the text.}
\label{fig:skew_demo}
\end{figure}

Fig.~\ref{fig:StirlingGroundTruth} demonstrates the accuracy of the Stirling and Gaussian approximations for a very low $u=1000$ and $q=0.4$.  The low value of $u$ and high value of $q$ were necessary in order to keep the graph from looking like a delta function, which is what a zoomed-out pdf of larger, more practically realistic, values of $u$ appears to be.  Even at such a low value of $u$, the simulation results, exact analytic pdf (Equation~\ref{eqn:ugly_SS}), Stirling approximation of the pdf (Equation~\ref{eqn:Stirling_approx}), and the Gaussian approximation of the pdf (Equation~\ref{eqn:Gaussian_approx}) are indistinguishable.  In order to see the slight skew predicted by higher order terms, we are forced to use an unrealistically low value of $u=25$, as seen in Fig.~\ref{fig:skew_demo}.  However, even in this unrealistically small example, both the Stirling and Gaussian approximations are close matches to the exact pdf. 

\begin{figure}
\centerline{\includegraphics{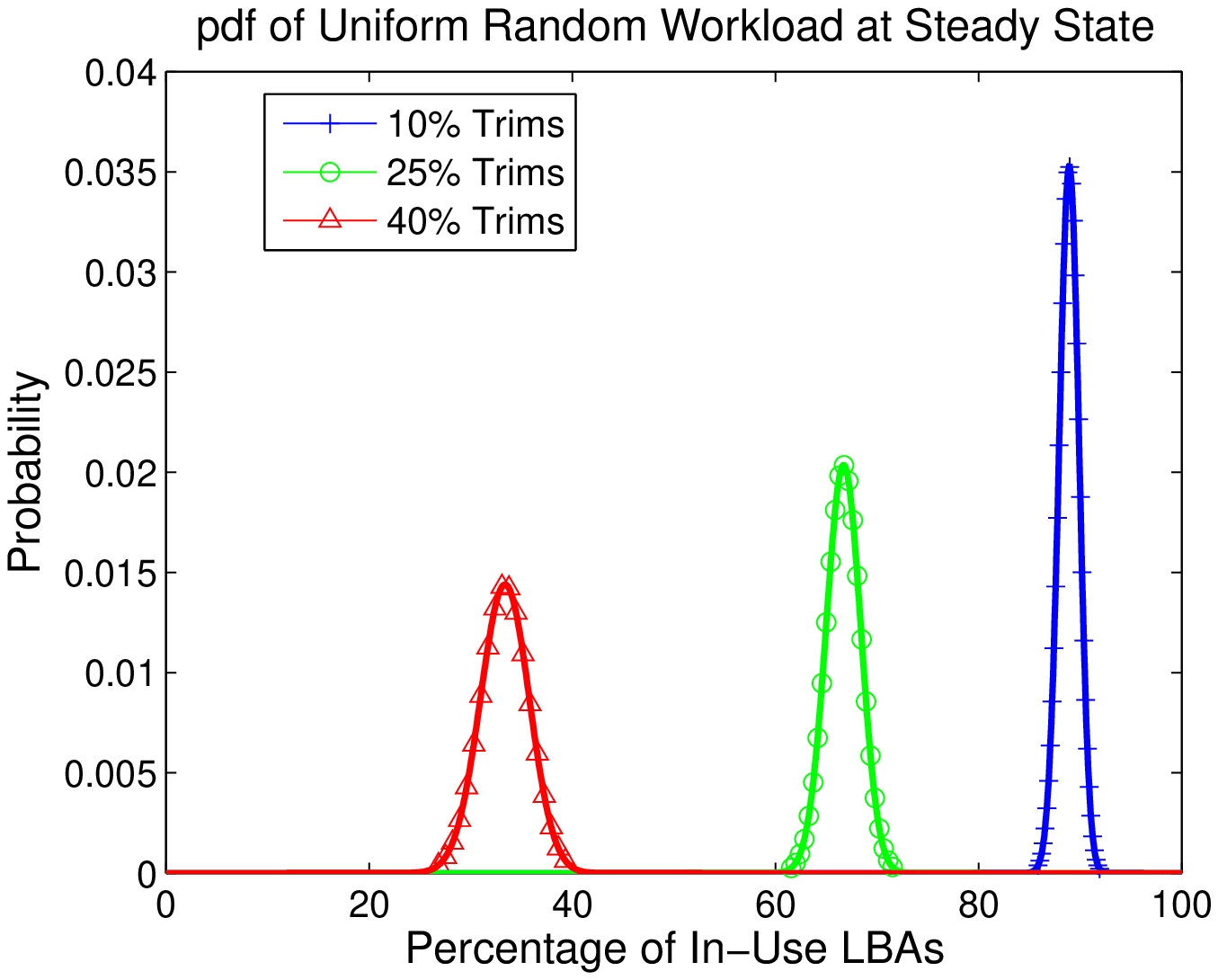}}
\caption{Theoretical results for utilization under various amounts of Trim. When more Trim requests are in the workload, a lower percentage of LBAs are In-Use at steady state, but the variance is higher than seen in a workload with fewer Trim requests.}
\label{fig:pdf_steady}
\end{figure}

\begin{figure}
\centerline{\includegraphics{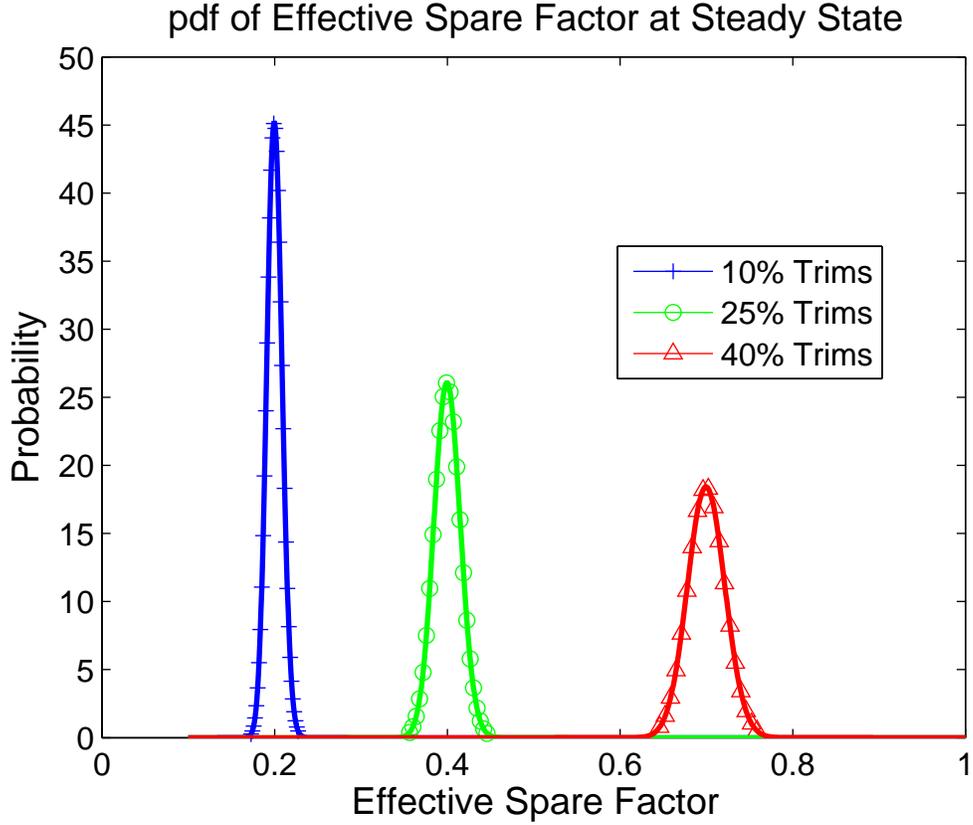}}
\caption{Theoretical results for the effective spare factor under various amounts of Trim, with a manufacturer specified spare factor of 0.11. When more Trim requests are in the workload, a higher effective spare factor is seen, but with a higher variance than seen in a workload with fewer Trim requests.}
\label{fig:pdf_spare_factor}
\end{figure}

\begin{figure}
\centerline{\includegraphics{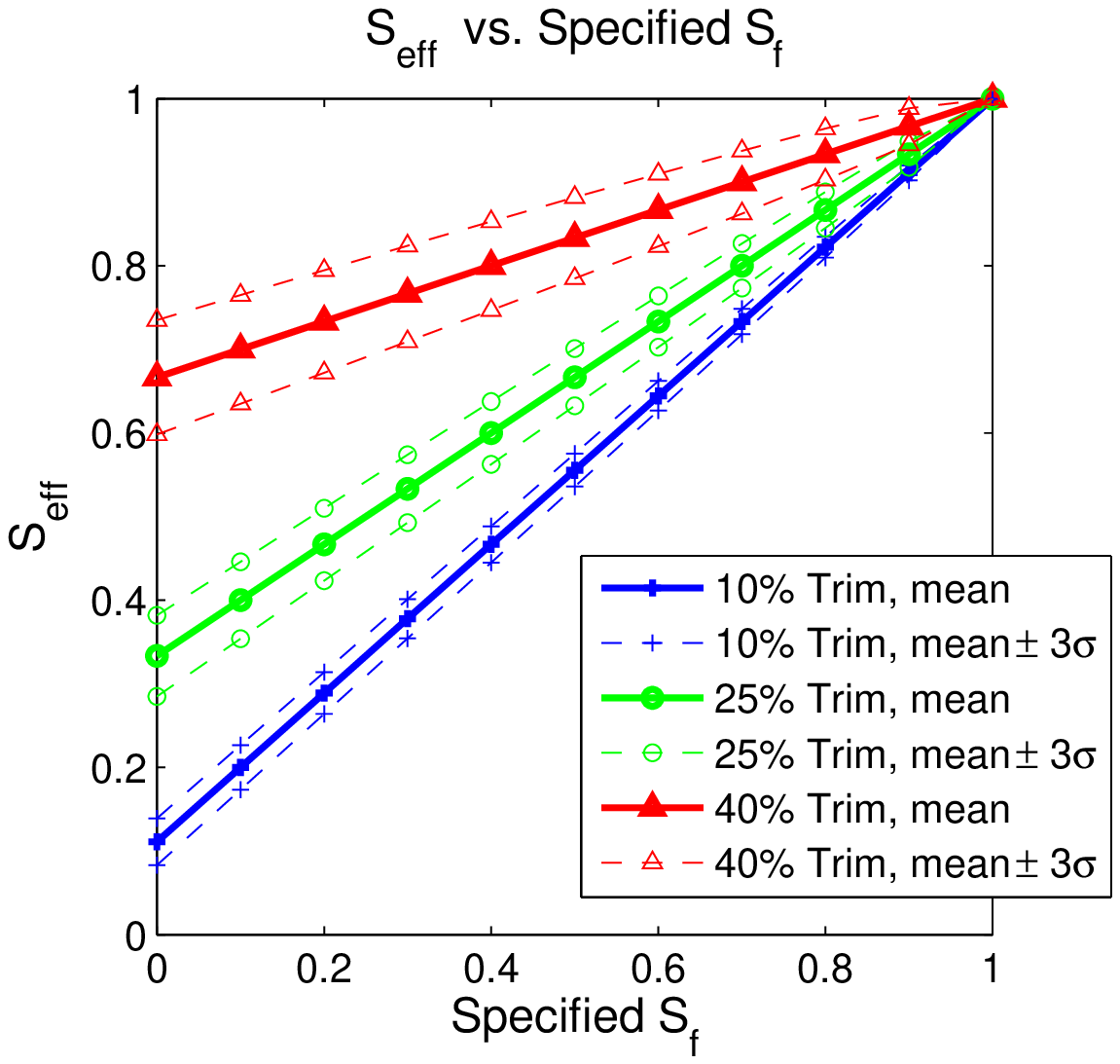}}
\caption{Theoretical results comparing the effective spare factor to the manufacturer-specified spare factor.  The mean and $3\sigma$ standard deviations are shown.  For this figure, a small $t=1280$ value was used so that the $3\sigma$ values are visible.  Larger gains in the effective spare factor are seen for smaller specified spare factor.  In this figure, we see that a workload of 10\% Trim requests transforms an SSD with a manufacturer-specified spare factor of zero into one with an effective spare factor of 0.11.  To obtain this spare factor through physical overprovisioning alone would require 12.5\% more physical space than the user is allowed to access!}
\label{fig:s_eff_vs_s_f}
\end{figure}

Fig.~\ref{fig:pdf_steady} illustrates the utilization of the user space as the percentage of In-Use LBAs at steady state for several rates of Trim.  Higher rates of Trim correspond to lower utilization, but with a wide variance.  The corresponding graph of the pdf of the steady state effective spare factor is found in Fig.~\ref{fig:pdf_spare_factor}, with the example utilizing a manufacturer specified spare factor of $0.11$.  Here, higher rates of Trim correspond to a higher effective spare factor, but with a wide variance.  Fig.~\ref{fig:s_eff_vs_s_f} relates the manufacturer specified spare factor to the effective spare factor, showing both the mean and $3\sigma$ standard deviations.  In this figure, we see that a workload of 10\% Trim requests transforms an SSD with a manufacturer-specified spare factor of zero into one with an effective spare factor of 0.11.  To obtain this spare factor through physical overprovisioning alone would require 12.5\% more physical space than the user is allowed to access! Implementing the Trim command on an SSD allows for comparable performance without requiring extra physical materials, which lowers the price of making the SSD.  Understanding the relationship between the workload and the performance of the device is necessary for consumers to be able to accurately compare SSD products; the theory in this paper sets out a framework to model this relationship.

\section{The Effect of Trim on Write Amplification}
\citeN{Hu2009} and~\citeN{hu2010fundamental},~\citeN{agarwal2010closed}, and~\citeN{xiang2011improved} have each derived an analytic approach for computing the write amplification under the standard uniform random workload.  We have previously published~\cite{frankie2012_ciit} an analysis of the Trim-modified uniform random workload that parallels the derivation by~\citeN{xiang2011improved}.

\subsection{Analysis of the Standard Uniform Random Workload}
\subsubsection{Hu's Approach}
\begin{figure}[t!]
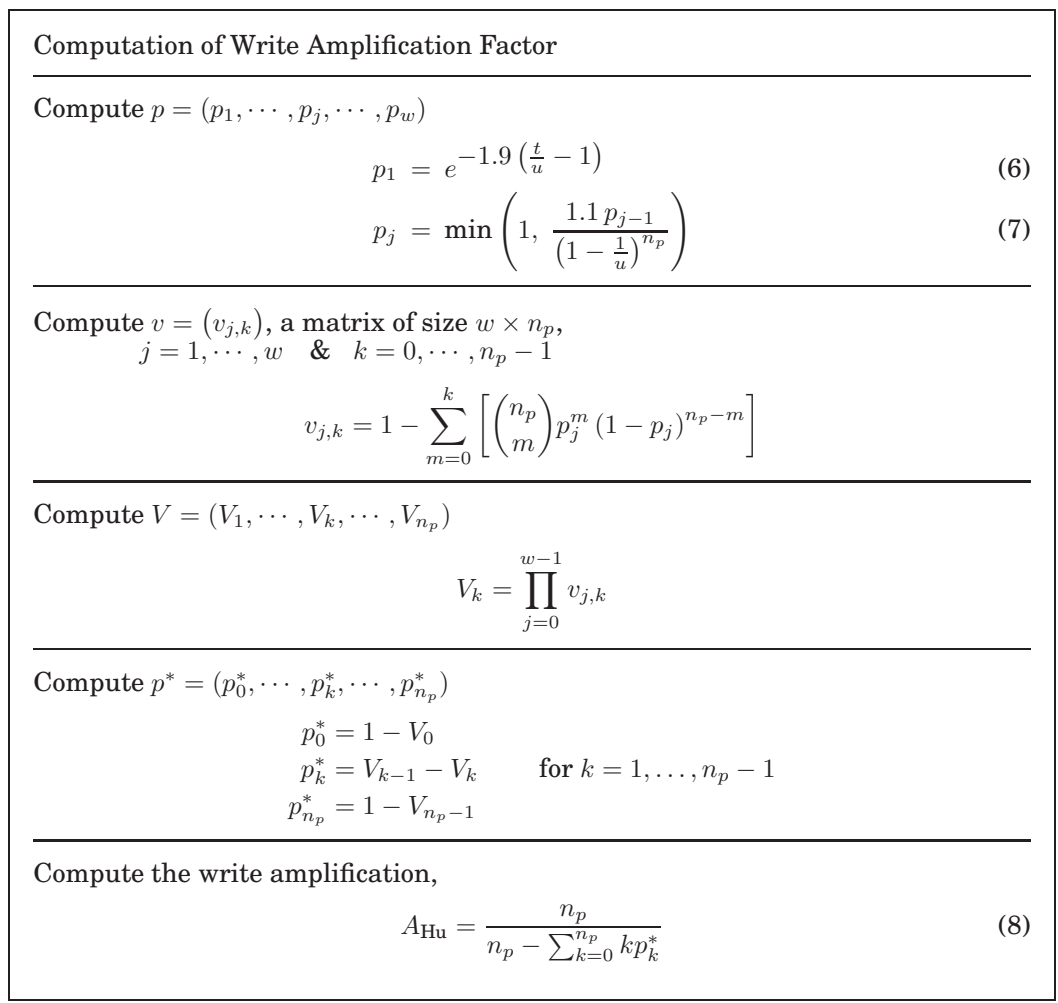

\begin{framed}
Computation of Write Amplification Factor
\medskip
\hrule \vspace{8pt}
Compute $p = (p_1, \cdots, p_j, \cdots, p_w)$
\begin{eqnarray}
p_1 &=& e^{\text{\normalsize$-1.9\left(\frac{t}{u}-1\right)$}} \label{eqn:Hu_needs_modify} \\
p_j &=& \text{min}\left(1, \; \frac{1.1 \, p_{j-1}}{ \left(1-\frac{1}{u}\right)^{n_p}} \right) \label{eqn:Hu_needs_modify2}
\end{eqnarray}
\hrule \vspace{8pt}
Compute $v = \big(v_{j,k}\big)$, a matrix of size $w \times n_p$, \newline \phantom{1} \hspace{10.2mm} $j= 1, \cdots, w$\quad \& \ \  $k= 0, \cdots, n_p-1$
\begin{equation*}
v_{j,k} = 1 - \sum_{m=0}^{k}\left[{n_p \choose m} p_j^m \left(1-p_j\right)^{n_p-m}\right]
\end{equation*}
\hrule \vspace{8pt}
Compute $V = (V_1, \cdots, V_k, \cdots, V_{n_p})$
\begin{equation*}
V_{k} = \prod_{j=0}^{w-1}v_{j,k}
\end{equation*}
\hrule \vspace{8pt}
Compute $p^* = (p^*_0, \cdots, p^*_k, \cdots, p^*_{n_p})$
\begin{align*}
p_0^* &= 1-V_{0} \\
p_k^* &= V_{k-1} - V_{k}  \qquad  \text{ for } k=1, \ldots, n_p-1 \\
p_{n_p}^* &= 1-V_{n_p-1}
\end{align*}
\hrule \vspace{8pt}
Compute the write amplification,
\begin{equation}\label{eqn:Hu_write_amp}
A_\text{Hu} = \frac{n_p}{n_p-\sum_{k=0}^{n_p} kp_k^*}
\end{equation}
\end{framed}
\caption{Equations needed to compute the empirical model-based algorithm for $A_\text{Hu}$ (adapted from~\cite{Hu2009,hu2010fundamental} and~\cite{frankie2012_ciit}).  $w$ is the window size, and allows for the windowed greedy garbage collection variation of greedy garbage collection.  Setting $w=\frac{t}{n_p}-r$ is needed for the standard greedy garbage collection discussed in this paper.  For an explanation of the theory behind these formulas, see~\cite{Hu2009,hu2010fundamental}.}
\label{eqn:fig_Hu}
\end{figure}

\citeN{Hu2009} and~\citeN{hu2010fundamental} utilize an empirical model to compute the write amplification of a purely random write workload under greedy garbage collection, noting that the value of parameters other than the device utilization do not substantially effect the write amplification.  The equations needed to compute write amplification under their model are found in Fig.~\ref{eqn:fig_Hu}.

\subsubsection{Agarwal's Approach}
\citeN{agarwal2010closed} probabilistically model the write amplification of a purely random write workload under greedy garbage collection.  They observed that the distribution of the number of blocks with $v$ valid pages can be approximated as a uniform distribution, then equate the expected value of this uniform distribution with the expected value of the binomially distributed number of valid pages in a block chosen for garbage collection\footnote{The number of valid pages in a block chosen for garbage collection is binomially distributed due to the random nature of the write requests.}.  The resulting equation predicts write amplification as
\begin{equation}\label{eqn:Agar_write_amp}
A_\text{Agarwal} = \frac{1}{2} \left( \frac{1+\rho}{\rho} \right)
\end{equation}
where $\rho = (t-u)/u$ is a measure of the level of overprovisioning.
\subsubsection{Xiang's Approach}
\citeN{xiang2011improved} improve the approach taken by~\citeN{agarwal2010closed}.  They probabilistically analyze the number of invalid pages freed by garbage collection, assuming that it reaches a steady state with constant expected value.  Additionally, they compute the probability of a page being invalid in a block chosen for garbage collection, then utilize the fact that all pages in the block are equally likely to be invalid to compute the binomially distributed expected number of invalid pages to be freed by garbage collection.  Equating these values, then taking the limit as the device becomes large, they find the write amplification is
\begin{equation}\label{eqn:Xiang_write_amp}
A_\text{Xiang} = \frac{-1-\rho}{-1-\rho-W((-1-\rho)e^{(-1-\rho)})}
\end{equation}
where $W(\cdot)$ denotes the Lambert W function~\cite{corless1996lambertw,xiang2011improved}.

\subsection{Analysis of the Trim-Modified Uniform Random Workload}
\subsubsection{Paralleling Xiang's Approach}
In~\cite{frankie2012_ciit}, we added Trim commands to the purely random write workload analyzed by~\citeN{Hu2009} and~\citeN{hu2010fundamental}, ~\citeN{agarwal2010closed}, and ~\citeN{xiang2011improved}, then paralleled the derivation of~\citeN{xiang2011improved} to compute the write amplification under greedy garbage collection, finding\footnote{We denote the calculation of write amplification with TRIM requests in the workload as $A^{\text{\tiny ($T$)}}$.}
\begin{equation}\label{eqn:Xiang_derived}
A_\text{Xiang}^{\text{\tiny ($T$)}} = \frac{\text{\large${\frac{-(1+\rho)}{s}}$}} {\text{\large${\frac{-(1+\rho)}{s}}$} - W\left(\frac{-(1+\rho)}{s} e^\text{\large${\frac{-(1+\rho)}{s}}$}\right)}
\end{equation}

\subsubsection{Modification of Equations for Hu and Agarwal Approaches}
A comparison of $A_\text{Xiang}^{\text{\tiny ($T$)}}$ and $A_\text{Xiang}$ reveals that they are the same equation, but with
\begin{equation}\label{eqn:eff_rho}
\rho_\text{eff} = \frac{\bar{S}_\text{eff}}{1-\bar{S}_\text{eff}} = \frac{1+\rho}{s}-1
\end{equation}
in place of $\rho$~\cite{frankie2012_ciit}.  This value for $\rho_\text{eff}$ can be placed into $A_\text{Agarwal}$ to obtain
\begin{equation*}
A_\text{Agarwal}^{\text{\tiny ($T$)}} = \frac{1}{2} \left( \frac{1+\rho}{1+\rho-s} \right)
\end{equation*}
as an expression accounting for Trim requests.  $A_\text{Hu}$ can also be modified by changing the equations in Fig.~\ref{eqn:fig_Hu} to
\begin{eqnarray*}
p_1 &=& e^{\text{\normalsize$-1.9\left(\frac{t}{u_\text{eff}}-1\right)$}} \\
p_j &=& \text{min}\left(1, \; \frac{1.1 \, p_{j-1}}{ \left(1-\frac{1}{u_\text{eff}}\right)^{n_p}} \right)
\end{eqnarray*}
where $u_\text{eff}=us$~\cite{frankie2012_ciit}.

\subsection{Simulation Results}
\begin{figure}
\centerline{\includegraphics{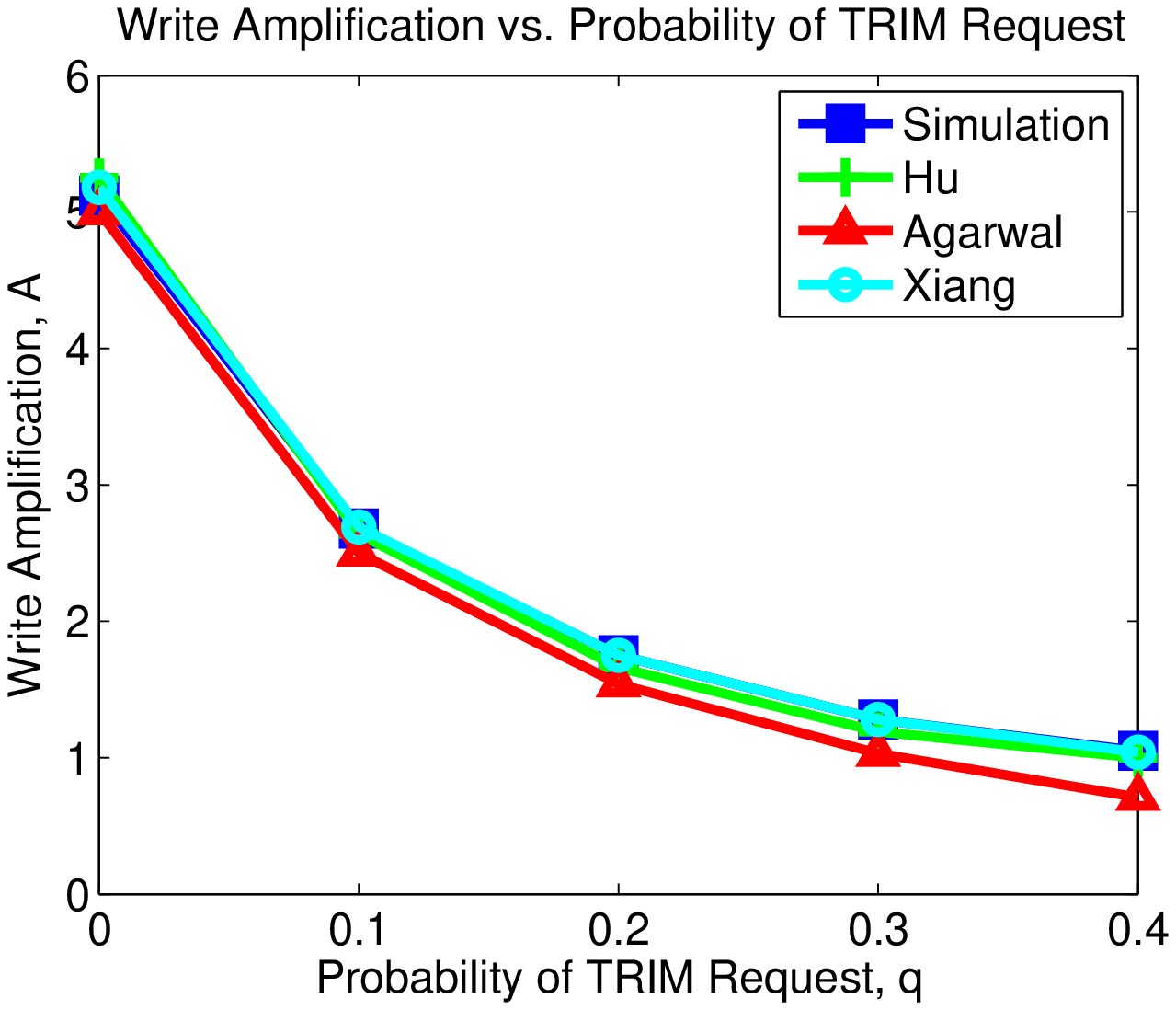}}
\caption{Simulation and theoretical results for write amplification under various amounts of Trim and $S_f=0.1$.  The Trim-modified Xiang calculation is a very good match.  The Trim-modified calculations for Hu and Agarwal are optimistic about the write amplification, with the Agarwal calculation being so optimistic as to give an impossible write amplification of less than 1 for high amounts of Trim!}
\label{fig:write_amp_sim_theory}
\end{figure}

A comparison of simulation and theoretical results is found in Fig.~\ref{fig:write_amp_sim_theory}.  $A_\text{Xiang}^{\text{\tiny ($T$)}}$ gives a very good approximation of the simulated write amplification values. $A_\text{Hu}^{\text{\tiny ($T$)}}$ and $A_\text{Agarwal}^{\text{\tiny ($T$)}}$ are good approximations for low probability of Trim requests, corresponding to low values of $\rho_\text{eff}$.  However, for larger values of $q$, corresponding to high values of $\rho_\text{eff}$, $A_\text{Hu}^{\text{\tiny ($T$)}}$and $A_\text{Agarwal}^{\text{\tiny ($T$)}}$ are optimistic in their predictions of write amplification, with $A_\text{Agarwal}^{\text{\tiny ($T$)}}$ becoming so optimistic that it predicts an impossible write amplification of less than 1.

\section{Hot and Cold Data}
The Trim-modified uniform random workload analyzed above is unlikely to be found exactly in real-world applications.  Most workloads will have some data that is accessed frequently (called hot data), and some data that is accessed infrequently (called cold data).  Rosenblum and Ousterhout discussed this type of data in their paper~\cite{rosenblum1992design}, suggesting that an appropriate mix of hot and cold data is that hot data is requested 90\% of the time, but only occupies 10\% of the user space, and cold data is requested 10\% of the time while occupying 90\% of the user space.  ~\citeN{Hu2009} and~\citeN{hu2010fundamental} showed through simulation that a standard uniform random workload consisting of hot and cold data under the log-structured writing and greedy garbage collection described above had high write amplification, but when the hot data and the cold data were kept in separate blocks, the write amplification improved.  In their simulation, the cold data was read-only.  We generalize the workload to allow cold data to be written and Trimmed infrequently compared to the hot data.  In this section, we perform analysis of a Trim-modified uniform random workload with hot and cold data in which the hot and cold data are kept separate, and use simulation to examine the resulting write amplification when hot and cold data are mixed together in blocks.  We assume that the temperature of the data is perfectly known; in reality, this information would need to be learned by the device; learning the temperature of the data is beyond the scope of this paper.

\subsection{Mixed Hot and Cold Data}
In the mixed hot and cold data scenario, no special treatment is given to the data based on its temperature; hot and cold data are mixed within blocks.  However, when all LBAs are not accessed uniformly, as with hot and cold data, the theory already described for computing write amplification does not hold.  We show this for the mixed case of hot and cold data by using the effective overprovisioning to compute the write amplification and comparing it with the simulation values.  To compute the effective overprovisioning, first compute the expected number of In-Use LBAs for hot and cold data, the add these to get $u_\text{eff}$.  Then, the effective overprovisioning is $\rho_\text{eff} = (t-u_\text{eff})/u_\text{eff}$ and can be plugged directly into the $\rho$ of Equation~\ref{eqn:Xiang_write_amp} to compute the (incorrect) write amplification.

\begin{figure}
\centerline{\includegraphics{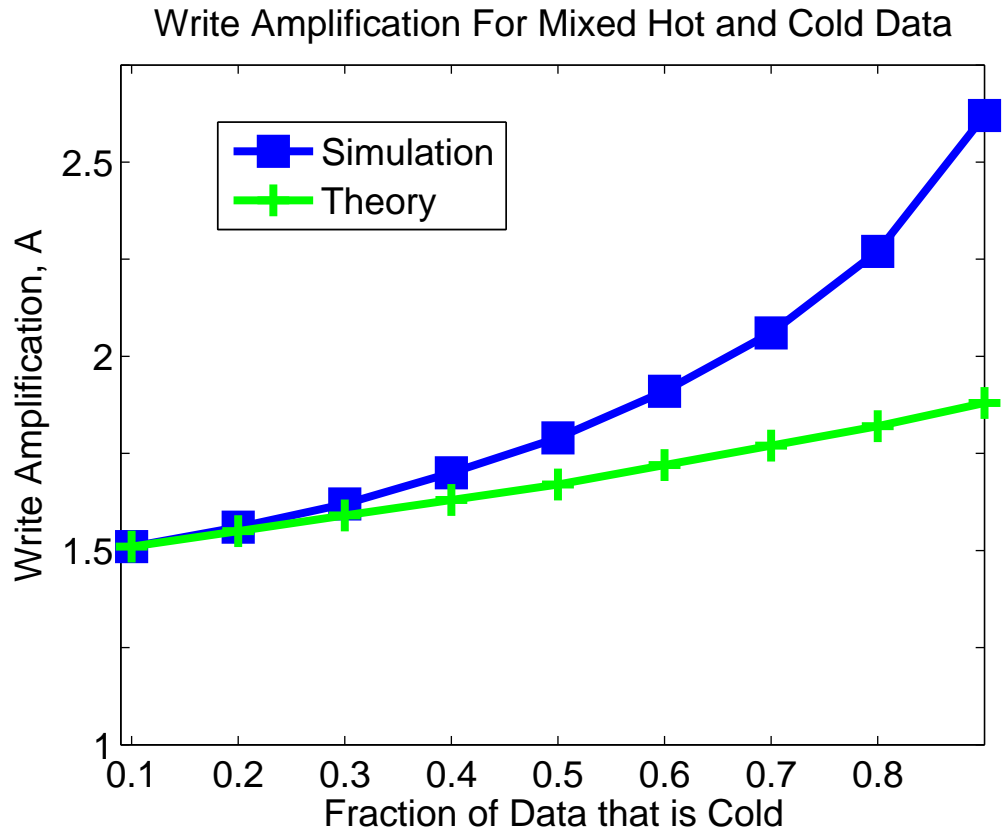}}
\caption{Write amplification for mixed hot and cold data.  Physical spare factor is 0.2.  Trim requests for cold data occur with probability 0.1; Trim requests for hot data occur with probability 0.2.  Cold data is requested 10\% of the time, with hot data requested the remaining 90\% of the time.  Notice that the theoretical values, based only on the level of effective overprovisioning, are optimistic when cold data accounts for more than 20\% of the user space.}
\label{fig:hot_cold_write_amp}
\end{figure}

As seen in Figure~\ref{fig:hot_cold_write_amp}, the theoretical results are not too far off when cold data is up to 20\% of the user space.  However, for larger fractions of cold data in the user space, the theoretical results are too optimistic to be of value, and the actual observed write amplification grows large. This is undesirable, since cold data is likely to occupy a larger fraction of the user space than hot data.

An analytic solution for the write amplification of mixed hot and cold data is not straightforward, and does not easily parallel the derivation of~\citeN{xiang2011improved}.  It is known that mixing hot and cold data leads to higher write amplification than keeping the data separate~\citeN{Hu2009} and~\citeN{hu2010fundamental}, so a mathematical analysis of the scenario is not likely to provide utility beyond mathematical interest.  Instead, we provide simulation results in Fig.~\ref{fig:hot_cold_write_amp} for comparison purposes with the separated data scenario.

\subsection{Separated Hot and Cold Data}\label{sec:separated_hot_cold}
Based on the results from~\citeN{Hu2009} and~\citeN{hu2010fundamental}, we expect to see an improvement in the write amplification over the mixed data case.  In the separated data scenario, a set of physical blocks are assigned to the cold data, and a separate set of physical blocks is assigned to the hot data.  Each set of blocks is kept logically separated~\footnote{The complete physical segregation of hot and cold blocks is useful for mathematical analysis.  However, in an actual device, wear leveling is an issue that needs to be addressed.  For this reason, physical blocks will occasionally need to be swapped between the hot and cold queues to spread the wear evenly.  If this is timed to coincide with near-simultaneous garbage collection of hot and cold blocks, the amount of write amplification contributed by this wear leveling is expected to be minimal.}, and data placement and garbage collection for each temperature operates independently, in the same way described in Algorithm~\ref{alg:greedy_garbage_collection}.

Assume that $h$ blocks are assigned to the hot data, and $c$ blocks are assigned to the cold data, where $h+c=t$.  Also assume that $f_h$ fraction of the data is hot, and $f_c$ fraction of the data is cold (note that $f_h + f_c = 1$), so that hot data has $u f_h$ LBAs and cold data has $u f_c$ LBAs.  We also allow the hot and cold data to have their own rates of Trim, $q_h$ and $q_c$, respectively\footnote{Allowing each temperature of data to have its own rate of Trim makes sense, as there is no reason to believe that hot and cold data will be Trimmed at the same rate.}.  Finally, the frequency at which requests arrive for hot and cold data\footnote{We expect requests for hot data to arrive more frequently than requests for cold data, hence the choice of names for each.  However, the analysis of this section does not require this to the be case.} is denoted as $p_h$ and $p_c$, respectively, where $p_h + p_c = 1$.  Although it may take longer for cold data to reach steady-state than it takes for hot data, we can compute the expected number of In-Use LBAs for hot and for cold data at steady-state.  We expect to see $u f_h s_h$ hot In-Use LBAs, and $u f_c s_c$ cold In-Use LBAs, with the number of valid pages the same as the number of In-Use LBAs for each temperature.  From this, we can compute the level of effective overprovisioning for each temperature, and then the write amplification.
\begin{equation}\label{eqn:write_amp_hot}
A_\text{Xiang}^{\text{\tiny ($T$, hot)}} = \frac{\text{\large${\frac{-(1+\rho_\text{hot})}{s_h}}$}} {\text{\large${\frac{-(1+\rho_\text{hot})}{s_h}}$} - W\left(\frac{-(1+\rho_\text{hot})}{s_h} e^\text{\large${\frac{-(1+\rho_\text{hot})}{s_h}}$}\right)}
\end{equation}
where $\rho_\text{hot} = \frac{h - u f_h}{u f_h}$.  A similar equation can be computed for $A_\text{Xiang}^{\text{\tiny ($T$, cold)}}$.  Finally, the two values are combined in a weighted fashion to obtain the overall write amplification.  Although it may seem the obvious choice, the weighting to use is not the ratio of hot and cold \emph{requests} to the total number of \emph{requests}; it is necessary to compute a ratio of hot and cold \emph{writes} to the total number of \emph{writes}
\begin{eqnarray}
\alpha_h &=& \frac{p_h(1-q_h)}{p_h(1-q_h) + p_c(1-q_c)} \\
\alpha_c &=& \frac{p_c(1-q_c)}{p_h(1-q_h) + p_c(1-q_c)}
\end{eqnarray}
\begin{equation}
A_\text{separated} = \alpha_h A_\text{Xiang}^{\text{\tiny ($T$, hot)}} + \alpha_c A_\text{Xiang}^{\text{\tiny ($T$, cold)}}
\end{equation}

\begin{figure}
\centerline{\includegraphics{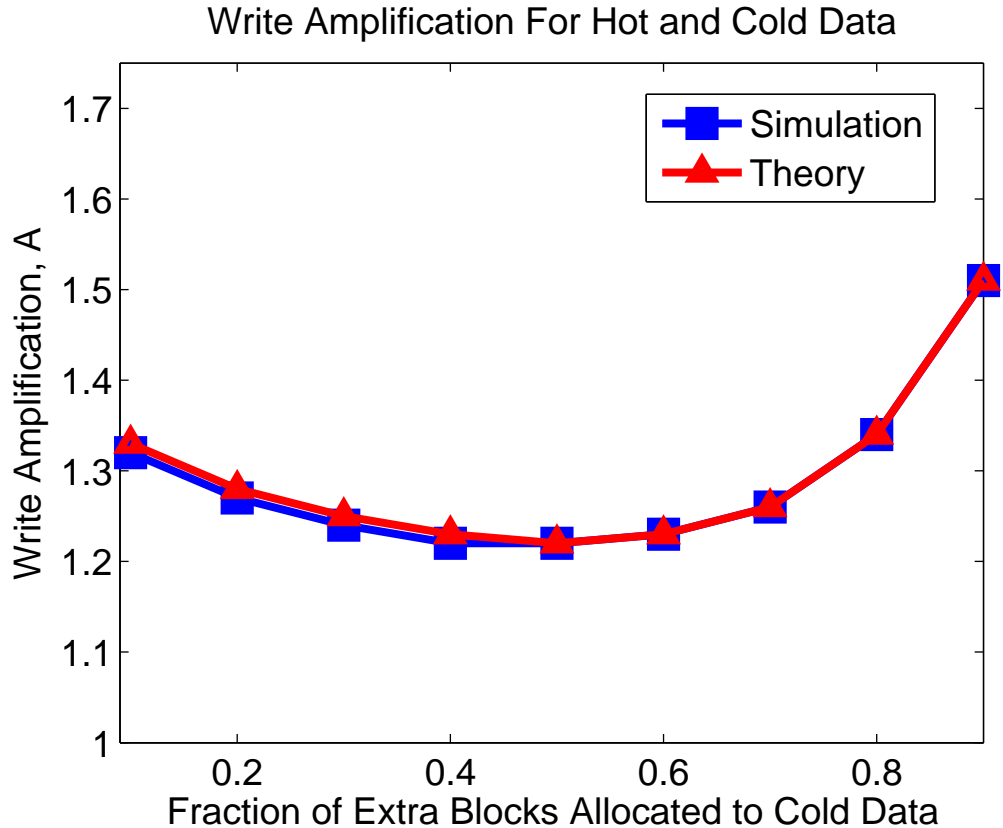}}
\caption{Write amplification for separated hot and cold data.  Physical spare factor is 0.2.  Trim requests for cold data occur with probability 0.1; Trim requests for hot data occur with probability 0.2.  Cold data is requested 10\% of the time, with hot data requested the remaining 90\% of the time.  After allocating the minimum number of blocks needed to store all of the user-allowed data for each temperature of data, the remaining blocks were split with the fraction shown on the x-axis.  In this case, the optimal split was 50-50; further investigation of the optimization problem is needed to determine whether this is always the case.}
\label{fig:hot_cold_spare_block_allocation}
\end{figure}

This analysis points to an optimization problem in how to allocate the overprovisioned physical blocks on the device after the minimum number of blocks needed for each temperature is assigned.  For the parameters chosen in graphing the results (see Fig.~\ref{fig:hot_cold_spare_block_allocation}), an equal allocation of the overprovisioned physical blocks is optimal.  However, it would require further exploration to determine whether this is true for all choices of parameters of if it is dependent on the parameter values.

\subsection{Simulation Results}
\begin{figure}
\centerline{\includegraphics{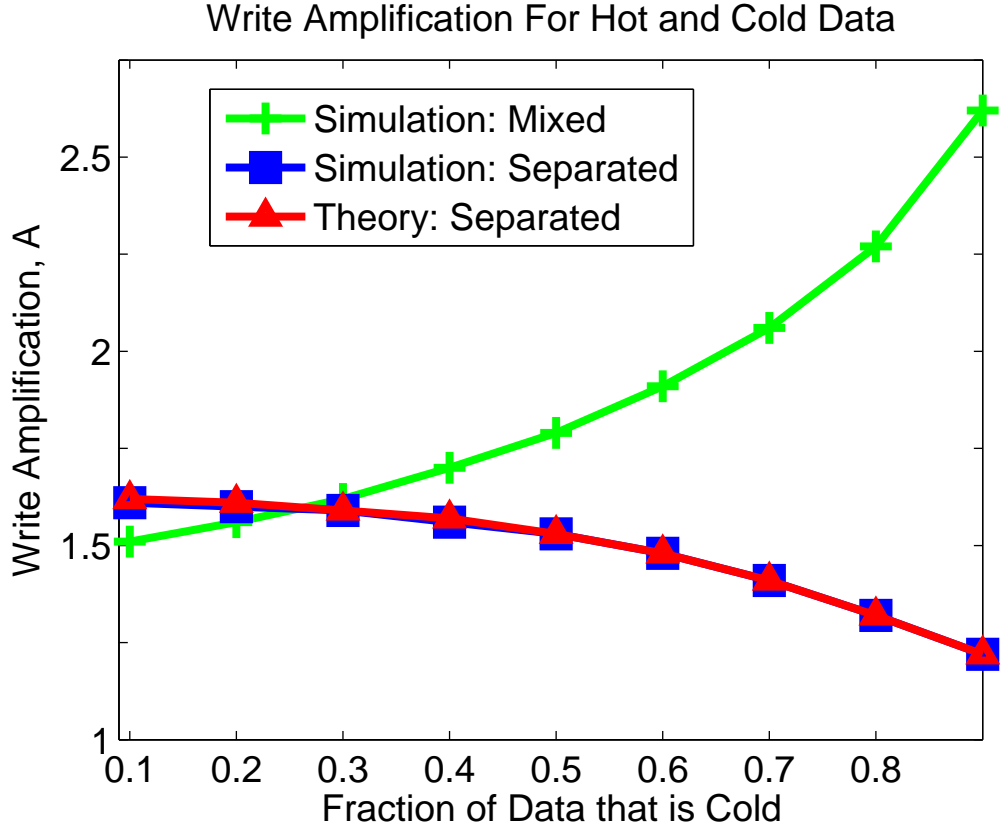}}
\caption{Write amplification for mixed hot and cold data compared to separated hot and cold data.  Physical spare factor is 0.2.  Trim requests for cold data occur with probability 0.1; Trim requests for hot data occur with probability 0.2.  Cold data is requested 10\% of the time, with hot data requested the remaining 90\% of the time.  For the separated hot and cold data, extra blocks were allocated evenly between the two data types.  The theoretical values for separated data are a good prediction of the simulation values.  Note that the separated hot and cold data has superior write amplification to the mixed hot and cold data.}
\label{fig:hot_cold_compare_mixed_separated}
\end{figure}

An examination of Fig.~\ref{fig:hot_cold_compare_mixed_separated} shows that physically separating hot and cold data reduces the write amplification over allowing the data to be mixed within blocks, unless cold data occupies only a very small fraction of the user space.  However, it is generally expected that infrequently accessed (cold) data, will be a large fraction of the data stored, so accurate determination of hot versus cold data in order to keep them on physically separate blocks is very important.

\section{$N$ Temperatures of Data}
The idea of hot and cold data can be extended to $N$ temperatures of data.  This can be used to model a more complicated workload consisting of $N$ temperatures of data, each with its own profile for probability of Trim.  Additionally, the idea of $N$ temperatures of data can also be thought of as many users who access data at different rates and with different amount of Trim in their workloads, which could be helpful in analyzing cloud storage workloads.  If each temperature of data is kept separate, the analysis of separated hot and cold data is easily extended to the case of $N$ temperatures.

Assume there are $N$ temperatures, with each temperature $j\in[1,N]$ having access to $u_j$ LBAs.  Then, the total number of user LBAs is $u=u_1+u_2+\cdots+u_N=\sum_{j=1}^{N} u_j$.  Also assume that each temperature has its own probability of Trim $q_j$ (and associated $s_j$), that is constant over time.  Then at steady state, the number of In-Use LBAs from user $j$ is $X_j$ with a mean of $u_j s_j$ and a variance of $u_j \bar{s}_j$.

Define the random variable $Y$ to be the total number of In-Use LBAs.  Then $Y = \sum_{j=1}^N X_j$.  Because $X_j$ are independent Gaussian random variables, their sum, $Y$, is also Gaussian, with mean $\sum_{j=1}^N u_j s_j$ and variance $\sum_{j=1}^N u_j \bar{s}_j$.  This information can be used to compute the overall effective overprovisioning.  However, as demonstrated in the hot and cold data section, the overall effective overprovisioning is not useful in computing the write amplification of a non-uniform workload.

The write amplification for N temperatures of data can be computed if the data is kept separated, as described in Section~\ref{sec:separated_hot_cold}~\footnote{Wear leveling would still need to be taken into consideration in a real device.}.  First, assume each temperature $j$ has $k_j$ space available for writing (so that $\sum_{j=1}^N k_j=t$) and frequency of writing $p_j$ (so that $\sum_{j=1}^N p_j=1$).  Then compute the value of $\rho_j = \frac{k_j - u_j}{u_j}$ for each temperature $j$, and use this to compute the write amplification for each temperature
\begin{equation*}
A_\text{Xiang}^{\text{\tiny ($T$, j)}} = \frac{\text{\large${\frac{-(1+\rho_j)}{s_j}}$}} {\text{\large${\frac{-(1+\rho_j)}{s_j}}$} - W\left(\frac{-(1+\rho_j)}{s_j} e^\text{\large${\frac{-(1+\rho_j)}{s_j}}$}\right)}
\end{equation*}
At first glance, it may appear that the overall write amplification is just a weighted sum of the individual write amplifications, where the weight is $p_j$.  If there are no Trim requests, this would be true.  However, the Trim requests in the workload need to be accounted for.  To properly weight the individual temperature write amplifications, we need a measure of the ratio of writes requested by each user.  Compute the weights $\alpha_j = \frac{p_j(1-q_j)}{\sum_{j=1}^N p_j(1-q_j)}$
Then, the overall write amplification is
\begin{equation*}
A_\text{separated, N} = \sum_{j=1}^N \alpha_j A_\text{Xiang}^{\text{\tiny ($T$, j)}}
\end{equation*}

%

\section{Conclusion}
In this paper, a comprehensive model has been presented for the Trim command in a uniform random workload.  The steady state number of In-Use LBAs was shown to be well approximated as a Gaussian random variable, but the analysis allows for inclusion of higher-order terms in order to compute higher order non-Gaussian corrections to moments such as skewness and kurtosis.  The steady-state number of In-Use LBAs was used to compute the level of effective overprovisioning, which allowed for the adaptation of previous non-Trim models to compute write amplification under a log-structured file system utilizing greedy garbage collection.

The Trim-modified uniform random workload was further extended to include various temperatures of data, allowing for closer approximation of real-world workloads.  The write amplification under this workload with up to $N$ temperatures of data was analytically computed under the condition that each temperature of data is stored in a physically segregated fashion.  For the case of $N=2$, hot and cold data, the importance of separating different temperatures of data in order to reduce write amplification was demonstrated.  Additionally, an optimization problem of how to allocate overprovisioned blocks to different temperatures of data was recognized, with the simulation result solution for one set of parameters shown.  Further work is needed to determine the optimal solution for all parameters.

Although this model is comprehensive, it still has limitations because of the assumptions made.  In this paper we used the standard assumption that one LBA requires one physical page in order to store the data~\cite{Hu2009,hu2010fundamental,agarwal2010closed,xiang2011improved}, which is equivalent to saying that all files stored are the same size.  However, in a real system, files are likely to be of varying sizes, perhaps requiring even more than one physical block in order to be stored; future work should incorporate this idea into the model.

\bibliographystyle{acmsmall}
\bibliography{collectedReferences}



\end{document}